\documentstyle[preprint,aps]{revtex}

\begin{document}
\title{{\bf Exact solutions of the pseudo-Coulomb potential plus ring-shaped
potential in the }$D${\bf -dimensional Schr\"{o}dinger equation by the
Nikiforov-Uvarov method }}
\author{Sameer M. Ikhdair\thanks{%
sikhdair@neu.edu.tr} and \ Ramazan Sever\thanks{%
sever@metu.edu.tr}}
\address{$^{\ast }$Department of Physics, \ Near East University, Nicosia, North
Cyprus, Mersin 10, Turkey\\
$^{\dagger }$Department of Physics, Middle East Technical University, 06531
Ankara, Turkey.}
\date{\today
}
\maketitle

\begin{abstract}
We present analytically the exact energy bound-states solutions of the
Schr\"{o}dinger equation in $D$-dimensions for an alternative (often used)
pseudo-Coulomb potential-plus- ring-shaped potential of the form $V(r)=-%
\frac{a}{r}+\frac{b}{r^{2}}+\frac{\beta \cos ^{2}\theta }{r^{2}\sin
^{2}\theta }+c$ by means of the conventional Nikiforov-Uvarov method. We
give a clear recipe of how to obtain an explicit solution to the radial and
angular parts of the wave functions in terms of orthogonal polynomials. The
total energy of the system is different from the pseudo-Coulomb potential
because of\ the contribution of the angular part. The general results
obtained in this work can be reduced to the standard forms given in
literature.


\end{abstract}


\section{Introduction}

\noindent The important task of quantum mechanics is to find the exact
bound-states solution of the Schr\"{o}dinger equation for certain potentials
of physical interest. Generally speaking, there are a few main traditional
methods to study the exact solutions of quantum systems like the Coulomb,
the harmonic oscillator [1,2], the pseudoharmonic [3,4] and the Kratzer
[4,5] potentials. Additionally, in order to obtain the bound-states
solutions of central potentials, one has to resort to numerical techniques
or approximation schemes. For many of the quantum mechanical systems, most
popular approximation methods such as shifted $1/N$ expansion [6],
perturbation theory [7], path integral solution [8], algebraic methods with
the SUSYquantum mechanics method and the idea of shape invariance, further
closely with the factorization mrthod [9], exact quantization rule [10,11],
the conventional Nikiforov and Uvarov (NU) method\ [12-25]. Some of these
methods have drawbacks in applications. Although some other methods give
simple relations for the eigenvalues, however, they lead to a very
complicated relations for the eigenfunctions.

The study of exact solutions of the Schr\"{o}dinger equation for a class of
non-central potentials with a vector potential and a non-central scalar
potential is of considerable interest in quantum chemistry [26-35]. In
recent years, numerous studies [36-40] have been made in analyzing the bound
states of an electron in a Coulomb field with simultaneous presence of
Aharanov-Bohm (AB) [41] field, and/or a magnetic Dirac monopole [42], and
Aharanov-Bohm plus oscillator (ABO) systems. In most of these studies, the
eigenvalues and eigenfunctions are obtained by means of seperation of
variables in spherical or other orthogonal curvilinear coordinate systems.
The path integral for particles moving in non-central potentials is
evaluated to derive the energy spectrum of this system analytically [43]. In
addition, the idea of SUSY and shape invariance is also used to obtain exact
solutions of such non-central but seperable potentials [44]. Very recently,
the conventional NU method has been used to give a clear recipe of how to
obtain an explicit exact bound-states solutions for the energy eigenvalues
and their corresponding wave functions in terms of orthogonal polynomials
for a class of non-central potentials [45].

Recently, Chen and Dong [46] found a new ring-shaped potential and obtained
the exact solution of the Schr\"{o}dinger equation for the Coulomb potential
plus this new ring-shaped potential which has possible applications to
ring-shaped organic molecules like cyclic polyenes and benzene. Recently,
Cheng and Dai [47], proposed a new potential consisting from the modified
Kratzer's potential [48] plus the new proposed ring-shaped potential in
[46]. They have presented the energy eigenvalues for this proposed
exactly-solvable non-central potential in three dimensional $($i.e., $D=3)$%
-Schr\"{o}dinger equation through the NU method. The two quantum systems
solved by Refs [46,47] are closely relevant to each other as they deal with
a Coulombic field interaction except for a slight additional centrifugal
potential barrier acts as a repulsive core which is for any arbitrary
angular momentum $\ell $ prevents collapse of the system in any dimensional
space due to the slight perturbation to the original angular momentum
barrier. In a very recent work [49], we have given a clear recipe of how to
obtain analytically the exact energy eigenvalues and the corresponding wave
functions of the $D$-dimensional space Schr\"{o}dinger equation with the
proposed potential in [47] by means of the conventional NU method.

The conventional Nikiforov-Uvarov (${\rm NU}$) method [12], which received
much interest in recent years, has been introduced for solving
Schr\"{o}dinger equation [13-21], Klein-Gordon [22,23], Dirac [24] and
Salpeter [25] equations. The purpose of this work is to follow parallel
solution to [47] and give exact complete solutions to the $D$-dimensional
Schr\"{o}dinger equation with pseudo-Coulomb potential plus ring-shaped
potential, a Coulombic-like potential with an additional centrifugal
barrier, for any arbitrary $\ell ^{\prime }$-states using the conventional
Nikiforov-Uvarov method. For some appropriate choices of parameters, the
obtained expressions of energy eigenvalues and their corresponding wave
functions for this alternative form of the potential coincide with the $D$-$%
\dim $ensional results given in [49] and with the three-dimensions results
given in [47] in the limiting case of $D=3.$

This work is organized as follows: in section \ref{BC}, we shall briefly
introduce the basic concepts of the NU method. Section \ref{ES} is mainly
devoted to the exact solution of the Schr\"{o}dinger equation in $D$%
-dimensions for this quantum system by means of the ${\rm NU}$ method.
Finally, the relevant results are discussed in section \ref{RAC}.

\section{Basic Concepts of the Method}

\label{BC}The NU method is based on reducing the second-order differential
equation to a generalized equation of hypergeometric type [12]. In this
sense, the Schr\"{o}dinger equation, after employing an appropriate
coordinate transformation $s=s(r),$ transforms to the following form:
\begin{equation}
\psi _{n}^{\prime \prime }(s)+\frac{\widetilde{\tau }(s)}{\sigma (s)}\psi
_{n}^{\prime }(s)+\frac{\widetilde{\sigma }(s)}{\sigma ^{2}(s)}\psi
_{n}(s)=0,
\end{equation}
where $\sigma (s)$ and $\widetilde{\sigma }(s)$ are polynomials, at most of
second-degree, and $\widetilde{\tau }(s)$ is a first-degree polynomial.
Using a wave function, $\psi _{n}(s),$ of \ the simple ansatz:

\begin{equation}
\psi _{n}(s)=\phi _{n}(s)y_{n}(s),
\end{equation}
reduces (1) into an equation of a hypergeometric type

\begin{equation}
\sigma (s)y_{n}^{\prime \prime }(s)+\tau (s)y_{n}^{\prime }(s)+\lambda
y_{n}(s)=0,
\end{equation}
where

\begin{equation}
\sigma (s)=\pi (s)\frac{\phi (s)}{\phi ^{\prime }(s)},
\end{equation}

\begin{equation}
\tau (s)=\widetilde{\tau }(s)+2\pi (s),\text{ }\tau ^{\prime }(s)<0,
\end{equation}
and $\lambda $ is a parameter defined as
\begin{equation}
\lambda =\lambda _{n}=-n\tau ^{\prime }(s)-\frac{n\left( n-1\right) }{2}%
\sigma ^{\prime \prime }(s),\text{ \ \ \ \ \ \ }n=0,1,2,....
\end{equation}
The polynomial $\tau (s)$ with the parameter $s$ and prime factors show the
differentials at first degree be negative. It is worthwhile to note that $%
\lambda $ or $\lambda _{n}$ are obtained from a particular solution of the
form $y(s)=y_{n}(s)$ which is a polynomial of degree $n.$ Further, the other
part $y_{n}(s)$ of the wave function (2) is the hypergeometric-type function
whose polynomial solutions are given by Rodrigues relation

\begin{equation}
y_{n}(s)=\frac{B_{n}}{\rho (s)}\frac{d^{n}}{ds^{n}}\left[ \sigma ^{n}(s)\rho
(s)\right] ,
\end{equation}
where $B_{n}$ is the normalization constant and the weight function $\rho
(s) $ must satisfy the condition [12]

\begin{equation}
\frac{d}{ds}w(s)=\frac{\tau (s)}{\sigma (s)}w(s),\text{ }w(s)=\sigma (s)\rho
(s).
\end{equation}
The function $\pi $ and the parameter $\lambda $ are defined as

\begin{equation}
\pi (s)=\frac{\sigma ^{\prime }(s)-\widetilde{\tau }(s)}{2}\pm \sqrt{\left(
\frac{\sigma ^{\prime }(s)-\widetilde{\tau }(s)}{2}\right) ^{2}-\widetilde{%
\sigma }(s)+k\sigma (s)},
\end{equation}
\begin{equation}
\lambda =k+\pi ^{\prime }(s).
\end{equation}
In principle, since $\pi (s)$ has to be a polynomial of degree at most one,
the expression under the square root sign in (9) can be arranged to be the
square of a polynomial of first degree [12]. This is possible only if its
discriminant is zero. In this case, an equation for $k$ is obtained. After
solving this equation, the obtained values of $k$ are substituted in (9). In
addition, by comparing equations (6) and (10), we obtain the energy
eigenvalues.

\section{Exact solutions of the quantum system with the NU method}

\label{ES}

\subsection{Seperating variables of the Schr\"{o}dinger equation}

The pseudo-Coulomb potential plus ring-shaped potential in spherical
coordinates is defined as [47]

\begin{equation}
V(r,\theta )=-\frac{a}{r}+\frac{b}{r^{2}}+\beta \frac{\cos ^{2}\theta }{%
r^{2}\sin ^{2}\theta }+c,
\end{equation}
where $a,b,$ $c$ and $\beta $ are positive real constants. The potential in
(11) coincides with the potential introduced by Cheng-Dai [47] if the
parameters take their proper values and also turns to be the modified
Kratzer potential in the limiting case of $\beta =0$ [48]$.$ In fact the
energy spectrum for this potential can be obtained directly by considering
it as special case of the general non-central seperable potentials [45].

Our aim is to derive analytically the energy spectrum for a moving particle
in the presence of a potential (11) in a very simple way. The $D$%
-dimensional space Schr\"{o}dinger equation in spherical polar coordinates
written for potential (11) takes the form [1,6]

\[
-\frac{\hbar ^{2}}{2\mu }\left[ \frac{1}{r^{D-1}}\frac{\partial }{\partial r}%
\left( r^{D-1}\frac{\partial }{\partial r}\right) +\frac{1}{r^{2}}\left(
\frac{1}{\sin \theta }\frac{\partial }{\partial \theta }\left( \sin \theta
\frac{\partial }{\partial \theta }\right) +\frac{1}{\sin ^{2}\theta }\frac{%
\partial ^{2}}{\partial \varphi ^{2}}-\frac{2\mu \beta }{\hbar ^{2}}\frac{%
\cos ^{2}\theta }{\sin ^{2}\theta }\right) \right] \psi (r,\theta ,\varphi )
\]
\begin{equation}
+\left[ -\frac{a}{r}+\frac{b}{r^{2}}+c-E\right] \psi (r,\theta ,\varphi )=0,
\end{equation}
where $\mu =\frac{m_{1}m_{2}}{m_{1}+m_{2}}$ being the reduced mass of the
two particles and $\psi (r,\theta ,\varphi )$ being the total wave function
separated as follows

\begin{equation}
\psi _{n\ell m}(r,\theta ,\varphi )=R(r)Y_{\ell }^{m}(\theta ,\varphi ),%
\text{ }R(r)=r^{-(D-1)/2}g(r),\text{ }Y_{\ell }^{m}(\theta ,\varphi
)=H(\theta )\Phi (\varphi ).
\end{equation}
On substituting equation (13) into (12) leads to a set of second-order
differential equations:
\begin{equation}
\frac{d^{2}\Phi (\varphi )}{d\varphi ^{2}}+m^{2}\Phi (\varphi )=0,
\end{equation}

\begin{equation}
\left[ \frac{1}{\sin \theta }\frac{d}{d\theta }\left( \sin \theta \frac{d}{%
d\theta }\right) -\frac{m^{2}}{\sin ^{2}\theta }-\frac{2\mu \beta }{\hbar
^{2}}\frac{\cos ^{2}\theta }{\sin ^{2}\theta }+\ell (\ell +D-2)\right]
H(\theta )=0,
\end{equation}
\begin{equation}
\left[ \frac{1}{r^{D-1}}\frac{d}{dr}\left( r^{D-1}\frac{d}{dr}\right) -\frac{%
\ell (\ell +D-2)}{r^{2}}\right] R(r)+\frac{2\mu }{\hbar ^{2}}\left[ E+\frac{a%
}{r}-\frac{b}{r^{2}}-c\right] R(r)=0.
\end{equation}
The solution in (16) is periodic and must satisfy the period boundary
condition $\Phi (\varphi +2\pi )=\Phi (\varphi )$ from which we obtain
\begin{equation}
\Phi _{m}(\varphi )=\frac{1}{\sqrt{2\pi }}\exp (\pm im\varphi ),\text{ \ }%
m=0,1,2,.....
\end{equation}
Further, equation (15) representing the angular wave equation takes the
simple form
\begin{equation}
\frac{d^{2}H(\theta )}{d\theta ^{2}}+\frac{\cos \theta }{\sin \theta }\frac{%
dH(\theta )}{d\theta }+\left[ \ell (\ell +D-2)-\frac{m^{2}+(2\mu \beta
/\hbar ^{2})\cos ^{2}\theta }{\sin ^{2}\theta }\right] H(\theta )=0,
\end{equation}
which has already been solved before in our previous work [49]. We present
the final results for the angular part of the wave function as
\begin{equation}
H_{m^{\prime }}(\theta )=\sqrt{\frac{(2\ell ^{\prime }+1)(\ell ^{\prime
}-m^{\prime })!}{2(\ell ^{\prime }+m^{\prime })!}}\sin (\theta )^{m^{\prime
}}P_{n}^{(m^{\prime },m^{\prime })}(\cos \theta ),
\end{equation}
where
\begin{equation}
n=-\frac{(1+2m^{\prime })}{2}+\frac{1}{2}\sqrt{(2\ell ^{\prime
}+1)^{2}+4\ell ^{\prime }(D-3)},
\end{equation}
with
\begin{equation}
\ell ^{\prime }=-\frac{(D-2)}{2}+\frac{1}{2}\sqrt{(D-2)^{2}+4(n+\sqrt{%
m^{2}+2\mu \beta /\hbar ^{2}})(n+1+\sqrt{m^{2}+2\mu \beta /\hbar ^{2}})},
\end{equation}
and \
\begin{equation}
\text{\ }m^{\prime }{}^{2}=m^{2}+2\mu \beta /\hbar ^{2}.
\end{equation}
On the other hand, we are left to solve equation (16). After lengthy, but
straightforward, calculations, Equation (16), representing the radial wave
equation can be rewritten as [6]:
\begin{equation}
\frac{d^{2}g(r)}{dr^{2}}+\left[ \frac{2\mu }{\hbar ^{2}}(E-c)+\frac{2\mu a}{%
\hbar ^{2}}\frac{1}{r}-\frac{\widetilde{\nu }+(2\mu b/\hbar ^{2})}{r^{2}}%
\right] g(r)=0,
\end{equation}
where
\begin{equation}
\widetilde{\nu }=\frac{1}{4}(M-1)(M-3),\text{ }M=D+2\ell .
\end{equation}
The two particles in equation (18) interacting via Coulombic-like field have
a slight change in the angular momentum barrier acts as a repulsive core
which for any arbitrary $\ell $ prevents collapse of the system in any space
dimension due to the additional centrifugal potential barrier. Thus,
equation (23) has to be solved by using the NU method in the next subsection.

\subsection{The solutions of the radial equation}

The aim of this subsection is to solve the problem with a different radial
separation function $g(r)$ in any arbitrary dimensions. We now study the
bound-states (real) solution $E<c$ of equation (23). Letting
\begin{equation}
\varepsilon =\sqrt{-\frac{2\mu }{\hbar ^{2}}(E-c)},\text{ }\alpha =\frac{%
2\mu a}{\hbar ^{2}},\text{ }\gamma =\widetilde{\nu }+\frac{2\mu b}{\hbar ^{2}%
},
\end{equation}
and substituting these expressions in equation (23), one gets
\begin{equation}
\frac{d^{2}g(r)}{dr^{2}}+\left( \frac{-\varepsilon ^{2}r^{2}+\alpha r-\gamma
}{r^{2}}\right) g(r)=0.
\end{equation}
To apply the conventional NU-method, equation (26) is compared with (1) and
the following expressions are obtained

\begin{equation}
\widetilde{\tau }(r)=0,\text{ \ \ \ }\sigma (r)=r,\text{ \ \ }\widetilde{%
\sigma }(r)=-\varepsilon ^{2}r^{2}+\alpha r-\gamma .
\end{equation}
Substituting the above expressions into equation (9) gives

\begin{equation}
\pi (r)=\frac{1}{2}\pm \frac{1}{2}\sqrt{4\varepsilon ^{2}r^{2}+4(k-\alpha
)r+4\gamma +1}.
\end{equation}
According to this conventional method, the expression in the square root be
the square of a polynomial. Thus, the two roots $k$ can be readily obtained
as

\begin{equation}
k=\alpha \pm \varepsilon \sqrt{4\gamma +1}.
\end{equation}
In view of that, we arrive at the following four possible functions of $\pi
(r):$%
\begin{equation}
\pi (r)=\left\{
\begin{array}{cc}
\frac{1}{2}+\left[ \varepsilon r+\frac{1}{2}\sqrt{4\gamma +1}\right] & \text{%
\ for }k_{1}=\alpha +\varepsilon \sqrt{4\gamma +1}, \\
\frac{1}{2}-\left[ \varepsilon r+\frac{1}{2}\sqrt{4\gamma +1}\right] & \text{%
\ for }k_{1}=\alpha +\varepsilon \sqrt{4\gamma +1}, \\
\frac{1}{2}+\left[ \varepsilon r-\frac{1}{2}\sqrt{4\gamma +1}\right] & \text{%
\ for }k_{2}=\alpha -\varepsilon \sqrt{4\gamma +1}, \\
\frac{1}{2}-\left[ \varepsilon r-\frac{1}{2}\sqrt{4\gamma +1}\right] & \text{%
\ for }k_{2}=\alpha -\varepsilon \sqrt{4\gamma +1}.
\end{array}
\right.
\end{equation}
The correct value of $\pi (r)$ is chosen such that the function $\tau (r)$
given by equation (5) will have negative derivative [12]. So we can select
the physical values to be

\begin{equation}
k=\alpha -\varepsilon \sqrt{4\gamma +1}\text{ \ \ and \ \ }\pi (r)=\frac{1}{2%
}-\left[ \varepsilon r-\frac{1}{2}\sqrt{4\gamma +1}\right] ,
\end{equation}
which yield
\begin{equation}
\tau (r)=-2\varepsilon r+(1+\sqrt{4\gamma +1}).
\end{equation}
Using equations (6) and (10), the following expressions for $\lambda $ are
obtained, respectively,

\begin{equation}
\lambda =\lambda _{n}=2N\varepsilon ,\text{ }N=0,1,2,...,
\end{equation}
\begin{equation}
\lambda =\alpha -\varepsilon (1+\sqrt{4\gamma +1}).
\end{equation}
So we can obtain the energy eigenvalues as

\begin{equation}
E_{N}=c-\frac{2\mu a^{2}/\hbar ^{2}}{\left( 2N+1+\sqrt{(M-1)(M-3)+8\mu
b/\hbar ^{2}+1}\right) ^{2}},\text{ }
\end{equation}
where

\begin{equation}
(M-1)(M-3)=4\widetilde{\nu }=(D-2)^{2}+4\ell ^{\prime }(\ell ^{\prime
}+D-2)-8\mu \beta /\hbar ^{2}-1,
\end{equation}
with $\ell ^{\prime }$ defined in (21). Therefore, the final energy spectra
in equation (35) take the following Coulombic-like form [7]
\begin{equation}
E_{N^{\prime }}=c-\frac{\mu a^{2}/\hbar ^{2}}{2\left( N^{\prime }\right) ^{2}%
},\text{ }N^{\prime }=0,1,2,...
\end{equation}
where

\begin{equation}
N^{\prime }=N+L+1=\frac{1}{2}\left[ 2N+\sqrt{(D-2)^{2}+4\ell ^{\prime }(\ell
^{\prime }+D-2)+8\mu (b-\beta )/\hbar ^{2}}+1\right] ,
\end{equation}
is simply obtained by means of substituting equation (36) into (35).

(i) If $a=2D_{e}r_{e},$ $b=D_{e}r_{e}^{2},$ $c=D_{e}$ and $D=3$ (modified
Kratzer potential-plus-ring-shaped potential), the equation (35), with the
help of equation (36), is transformed into the following form

\begin{equation}
E_{Nnm}=D_{e}-\frac{8\mu D_{e}^{2}r_{e}^{2}/\hbar ^{2}}{\left( 2N+1+\sqrt{%
(2n+1)^{2}+4m^{2}+4(2n+1)\sqrt{m^{2}+2\mu \beta /\hbar ^{2}}+8\mu
D_{e}r_{e}^{2}/\hbar ^{2}}\right) ^{2}},
\end{equation}
and it is consistent with Ref. [47].

(ii) If $a=2D_{e}r_{e},$ $b=D_{e}r_{e}^{2},$ $c=D_{e},D=3$ and $\beta =0$
(modified Kratzer potential)$,$ the equation (35) is transformed into the
form

\begin{equation}
E_{n}=D_{e}-\frac{8\mu D_{e}^{2}r_{e}^{2}/\hbar ^{2}}{\left( 1+2n+\sqrt{%
1+4\ell (\ell +1)+8\mu b/\hbar ^{2}}\right) ^{2}}.
\end{equation}
and it is consistent with Ref. [48].

(iii) If $a=2D_{e}r_{e},$ $b=D_{e}r_{e}^{2}$ and $c=D_{e},$ the equation
(35) is transformed into the form

\begin{equation}
E_{Nnm}=D_{e}-\frac{8\mu D_{e}^{2}r_{e}^{2}/\hbar ^{2}}{\left( 2N+1+\sqrt{%
(D-2)^{2}+4\ell ^{\prime }(\ell ^{\prime }+D-2)+8\mu (D_{e}r_{e}^{2}-\beta
)/\hbar ^{2}}\right) ^{2}},
\end{equation}
and it is consistent with Ref. [49].

(iv) If $a=Ze^{2},$ $b=c=0$ and $D=3,$ the equation (35) is transformed into
the form

\begin{equation}
E_{Nnm}=-\frac{2\mu Z^{2}e^{4}/\hbar ^{2}}{\left( 2N+1+\sqrt{1+4\left[
\left( n+\sqrt{m^{2}+2\mu \beta /\hbar ^{2}}\right) \left( n+\sqrt{%
m^{2}+2\mu \beta /\hbar ^{2}}+1\right) -2\mu \beta /\hbar ^{2}\right] }%
\right) ^{2}},
\end{equation}
where $m,n,N=0,1,2,...$ and it is consistent with Ref. [46].

Let us now turn attention to find the radial wavefunctions for this
potential. Using $\tau (r),$ $\pi (r)$ and $\sigma (r)$ in equations (4) and
(8), we find
\begin{equation}
\phi (r)=r^{(\sqrt{4\gamma +1}+1)/2}e^{-\varepsilon r},
\end{equation}

\begin{equation}
\rho (r)=r^{\sqrt{4\gamma +1}}e^{-2\varepsilon r}.
\end{equation}
Then from equation (7), we obtain

\begin{equation}
y_{n}(r)=B_{n}r^{-\sqrt{4\gamma +1}}e^{2\varepsilon r}\frac{d^{N}}{dr^{N}}%
\left( r^{N+\sqrt{4\gamma +1}}e^{-2\varepsilon r}\right) ,
\end{equation}
and the wave function $g(r)$ can be written in the form of the generalized
Laguerre polynomials as

\begin{equation}
g(r)=C_{N,L}r^{L+1}e^{-\varepsilon r}L_{N}^{2L+1}(2\varepsilon r),
\end{equation}
where

\begin{equation}
L=\frac{1}{2}\left[ \sqrt{(D-2)^{2}+4\ell ^{\prime }(\ell ^{\prime
}+D-2)+8\mu (b-\beta )/\hbar ^{2}}-1\right] .
\end{equation}
Finally, the radial wave functions of the Schr\"{o}dinger equation are
obtained
\begin{equation}
R(r)=C_{N,L}r^{L-(D-3)/2}e^{-\varepsilon r}L_{N}^{2L+1}(2\varepsilon r),
\end{equation}
where

\begin{equation}
\varepsilon =\frac{\mu a}{\hbar ^{2}N^{\prime }}
\end{equation}
with $N^{\prime }$ is given in equation (38) and $C_{N,L}$ is the
normalization constant to be determined below. Using the normalization
condition, $\int\limits_{0}^{\infty }R^{2}(r)r^{D-1}dr=1,$ and the
orthogonality relation of the generalized Laguerre polynomials, $%
\int\limits_{0}^{\infty }z^{\eta +1}e^{-z}\left[ L_{n}^{\eta }(z)\right]
^{2}dz=\frac{(2n+\eta +1)(n+\eta )!}{n!},$ we have

\begin{equation}
C_{N,L}=\sqrt{\frac{(2\varepsilon )^{2L+3}N!}{2(N+L+1)(N+2L+1)!}}.
\end{equation}
Letting $D=3$ and $a=\alpha ,$ the equation (48) transforms into equation
(20) in Ref.[46]. Therefore, we may express the normalized total wave
functions as

\[
\psi (r,\theta ,\varphi )=\sqrt{\frac{(2\varepsilon )^{2L+3}(2\ell ^{\prime
}+1)(\ell ^{\prime }-m^{\prime })!N!}{4\pi (\ell ^{\prime }+m^{\prime
})!(N+L+1)(N+2L+1)!}}r^{L-(D-3)/2}\exp (-\varepsilon
r)L_{N}^{2L+1}(2\varepsilon r)
\]
\begin{equation}
\times \sin (\theta )^{m^{\prime }}P_{n}^{(m^{\prime },m^{\prime })}(\cos
\theta )\exp (\pm im\varphi ).
\end{equation}

\section{Results and Conclusions}

\label{RAC}In this paper, the Schr\"{o}dinger equation in any arbitrary
dimensions has been solved for its exact bound-states with a a
pseudo-Coulomb potential plus ring-shaped potential by means of a simple
conventional NU method. The analytical expressions for the total energy
levels of this system is found to be different from the results obtained for
the modified Kratzer's potential in [48] and also more general than the one
obtained recently in $D$-dimensions [49]. Hence, the noncentral potentials
treated in [45] can be introduced as perturbation to the central potentials
by adjusting the strength of the coupling constant $\beta $ in the
non-central part in terms of the coupling constants of the central part. In
addition, the angular part, the radial part and then the total wave
functions are also found. Thus, the Schr\"{o}dinger equation with a new
non-central but seperable potential has also been studied (cf. [45,49]and
the references therein). This method is very simple and useful in solving
other complicated systems analytically without given a restiction conditions
on the solution of some quantum systems as the case in the other models.
Finally, we point out that these exact results obtained for this new
proposed form of the potential (11) may have some interesting applications
in the study of different quantum mechanical systems, atomic and molecular
physics.

\acknowledgments This research was partially supported by the
Scientific and Technological Research Council of Turkey.     S.M.
Ikhdair wishes to dedicate this work to his family for their love
and assistance.

\end{document}